\documentclass{article}
\usepackage{spconf,amsmath,graphicx,hyperref}
\usepackage{algorithm}
\usepackage{algorithmic}
\usepackage{amsfonts}
\usepackage{url}
\usepackage{geometry}
\usepackage{graphicx}
\usepackage{float}
\usepackage{subfig}
\usepackage{multirow}
\usepackage{multicol}
\usepackage{mathtools}
\usepackage{booktabs}
\usepackage{subfig}
\usepackage{xcolor}
\usepackage{subfloat}
\title{Doppler Radiance Field-Guided Antenna Selection for Improved Generalization in Multi-Antenna Wi-Fi-based Human Activity Recognition}
\name{Navid Hasanzadeh{$^{1}$} and Shahrokh Valaee$^{1}$, Fellow, IEEE}
\address{$^{1}$Department of Electrical \& Computer Engineering, University of Toronto, Toronto, ON, Canada\\{\textit{\small  navid.hasanzadeh@mail.utoronto.ca, valaee@ece.utoronto.ca.}}}
\begin{document}
%\ninept
%
\maketitle
\begin{abstract}
With the IEEE 802.11bf Task Group introducing amendments to the WLAN standard for advanced sensing, interest in using Wi-Fi Channel State Information (CSI) for remote sensing has surged. Recent findings indicate that learning a unified three-dimensional motion representation through Doppler Radiance Fields (DoRFs) derived from CSI significantly improves the generalization capabilities of Wi-Fi-based human activity recognition (HAR). Despite this progress, CSI signals remain affected by asynchronous access point (AP) clocks and additive noise from environmental and hardware sources. Consequently, even with existing preprocessing techniques, both the CSI data and Doppler velocity projections used in DoRFs are still susceptible to noise and outliers, limiting HAR performance. To address this challenge, we propose a novel framework for multi-antenna APs to suppress noise and identify the most informative antennas based on DoRF fitting errors, which capture inconsistencies among Doppler velocity projections. Experimental results on a challenging small-scale hand gesture recognition dataset demonstrate that the proposed DoRF-guided Wi-Fi–based HAR approach significantly improves generalization capability, paving the way for robust real-world sensing deployments.
\end{abstract}
\section{Introduction}
\label{sec:intro}

Wi-Fi sensing is moving rapidly toward everyday use, powered by the IEEE $802.11$bf amendment that defines standard protocols for WLAN sensing and opens access to license-exempt bands below $7$ GHz~\cite{du2024overview}. By leveraging existing Wi-Fi infrastructure, eliminating the need for wearable devices, and mitigating the privacy concerns associated with cameras, Wi-Fi sensing provides a practical and accessible alternative to conventional sensing methods~\cite{radwan2025tutorial}. In the near future, Wi-Fi devices in homes, schools, hospitals, and workplaces could seamlessly perform human activity recognition (HAR), revolutionizing healthcare, smart environments, and immersive digital experiences.

Wi-Fi-based HAR analyzes variations in Channel State Information (CSI) caused by reflections and scattering to track motion~\cite{salehinejad2022litehar,ding2024multiple}. While prior studies have used either CSI magnitude or phase~\cite{yousefi2017survey,djogoHAR,jang2025study,zhang2024csi,peng2023rosefi} with machine learning techniques, these methods perform reliably only in static environments. Their sensitivity to environmental changes leads to limited accuracy and poor generalization across unseen users or locations~\cite{varga2024exposing}.

Recent methods extract Doppler velocity from Wi-Fi CSI to capture motion-induced frequency shifts while suppressing static components~\cite{meneghello2022sharp, zhang2021widar3}. MORIC~\cite{hasanzadeh2025moric} extends this idea by modeling Doppler projections as views from virtual cameras on a sphere, with each camera summarizing multipath effects through a von Mises–Fisher distribution. This structured representation improves generalization but remains limited by random and incomplete viewpoints. Inspired by neural radiance fields (NeRFs)~\cite{mildenhall2021nerf} in machine vision, Doppler radiance fields (DoRFs)~\cite{hasanzadeh2025dorf} address this issue by reconstructing a unified three-dimensional latent representation of motion from one-dimensional Doppler projections. DoRF integrates scattered views into a coherent global model, simulating observations from all viewpoints and enabling stronger generalization to unseen users and environments.

Despite the advances achieved by DoRF, noise in CSI signals can introduce spurious variations into Doppler projections and the resulting DoRF, leading machine learning models to misclassify activities in HAR, particularly when the activities are very similar. In fact, even minor misinformation on motion can markedly degrade performance in fine-grained tasks such as hand gesture recognition. Consequently, unlike NeRF in computer vision, where images used for three-dimensional reconstruction are typically less noisy, the high noise levels in CSI data severely limit the generalization capability of methods such as DoRF. Many of these noise sources distort the CSI phase in a highly non-linear manner, producing sudden phase shifts and spikes caused by multipath interference, local oscillator drift, and phase-locked loop (PLL) imperfections in one Wi-Fi antenna. As a result, traditional signal processing methods often fail to detect and suppress these artifacts effectively. Even advanced techniques like the CSI ratio model~\cite{wu2022wifi}, which attempts to cancel common noise across antenna pairs in a Wi-Fi access point (AP), fall short since some noise patterns occur exclusively in one antenna and not the others.

This work mitigates the impact of CSI noise in DoRF-based Wi-Fi HAR by using DoRF itself to refine motion representations. In the first stage, a DoRF model is trained on each AP independently, and antennas with high fitting errors are dynamically identified and discarded as noise sources, allowing the selection process to adapt automatically to varying noise levels and environmental conditions. In the second stage, DoRF models are trained on the remaining antennas, and their motion representations are used for activity classification. Experimental results show that the proposed two-stage DoRF-based HAR approach produces a motion representation that is more robust to noise and significantly improves generalization to unseen users, outperforming existing methods in challenging fine-grained human gesture recognition.

\section{Doppler Radiance Fields}
\label{sec:format}
\begin{figure*}[!t]
	\centering
	\subfloat{%
		\includegraphics[width=1.0\linewidth]{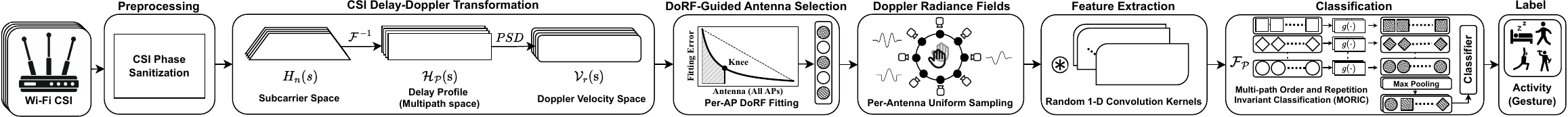}
	}
	\caption{The diagram summarizes the proposed Wi-Fi-based HAR framework. CSI is converted to Doppler velocity projections, which drive DoRF-guided antenna selection, retaining only antennas with consistent representation. The selected projections form DoRFs, which are uniformly sampled to obtain multiview representations. A feature extractor and a classifier then recognize the activity.
}
\label{figure:method}
\end{figure*}
This section summarizes the derivation of Doppler projections from Wi-Fi CSI and their use in constructing DoRFs as a compact informative representation of human motion.

For a Wi-Fi system with \(N\) orthogonal frequency-division multiplexing (OFDM) subcarriers spaced by \(\Delta f\) around a central carrier frequency \(f_c\), the channel response at subcarrier \(n\) and time \(s\) is modeled as  
\begin{equation}
H_n(s) = \sum_{l=0}^{L-1} \beta_l(s)\, 
e^{-j 2\pi \left(f_c - (n-\tfrac{N}{2})\Delta f\right)\tau_l(s)},
\end{equation}
where \(H_n(s)\) is the complex CSI, \(\beta_l(s)\) and \(\tau_l(s)\) denote the amplitude and delay of the \(l\)-th multipath component, and \(L\) is the total number of paths.  
The delay profile is obtained by applying an \(N\)-point inverse discrete Fourier transform (IDFT) across the subcarriers:
\begin{equation}
h(s,\tau)=\tfrac{1}{N}\sum_{n=0}^{N-1}H_n(s)e^{j2\pi n\Delta f\tau}.
\end{equation}
This transformation decomposes the CSI into delay bins \(\tau_i=i/(N\Delta f)\), where phase alignment causes energy at each \(\tau_i\) to add constructively, isolating the contributions of individual multipath components. Human motion perturbs the channel by introducing an extra delay to the \(i\)-th path:
\begin{equation}
\Delta \tau_i = \frac{\mathbf{v}(s)^\top \mathbf{m}_i}{c} t,
\end{equation}
with \(\mathbf{m}_i \in \mathbb{R}^3\) as the average direction of arrival for that path relative to the moving target, \(\mathbf{v}(s)\) the three-dimensional velocity vector over a brief interval around \(s\), \(c\) the propagation speed of electromagnetic waves, and \(t\) the time elapsed during motion. Consequently, the channel response at \(\tau_i\) experiences a temporal phase shift proportional to \(\mathbf{v}(s)^\top \mathbf{m}_i\), conceptualizing each \(\mathbf{m}_i\) as a virtual observation point similar to a 1D camera capturing motion projections from varied angles.

The power spectral density (PSD) for the delay bin \(\tau_i\) is obtained by taking the Fourier transform of its autocorrelation:  
\begin{equation}
S(f;\tau_i)=\int_{-\infty}^{\infty} 
\mathbb{E}\!\left[h^*(s;\tau_i)h(s+t;\tau_i)\right]
e^{-j2\pi ft}\,dt.
\end{equation}
This spectrum attains its peak at Doppler shift \(f^\star=\mathbf{v}(s)^\top \mathbf{m}_i/\lambda\) with wavelength \(\lambda=c/f_c\), giving the radial velocity
\begin{equation}
v_r(s;\tau_i)=\mathbf{v}(s)^\top \mathbf{m}_i.
\end{equation}
Collecting all radial velocities yields the Doppler field
\[
\mathcal{V}_r(s)=\{v_r(s;\tau_i)\}_{i=0}^{N-1},
\]
which represents multi-perspective projections of the true velocity. Let \(V_r \in \mathbb{R}^{T \times N}\) be the matrix compiling these Doppler velocities across \(T\) time instances, with entries \(V_r(s, i) = v_r(s; \tau_i)\). The velocity \(\mathbf{v}(s) \in \mathbb{R}^3\) serves as an approximation to the collective center-of-mass motion within the scene. Each projection satisfies
\[
v_r(s; \tau_i) = \mathbf{v}(s)^\top \mathbf{r}_i + n(s, i),
\]
where \(\mathbf{r}_i \in \mathbb{S}^2\) is an unidentified unit vector for direction, and \(n(s, i)\) accounts for observational noise. In matrix form, this becomes \(\mathbf{V}_r = \mathbf{V} \mathbf{R}^\top + \mathbf{N}\), with \(\mathbf{V} \in \mathbb{R}^{T \times 3}\) housing the velocities, \(\mathbf{R} \in \mathbb{R}^{N \times 3}\) the directions, and \(\mathbf{N}\) the noise matrix.

The inference of \(\mathbf{V}\) and \(\mathbf{R}\) is posed as a regularized optimization problem:
\begin{align}
\min_{\mathbf{V},\mathbf{R}}\quad &
\frac{1}{2} \sum_{s=0}^{T-1} \sum_{i=0}^{N-1}
\left[ \mathbf{V}_r(s,i) - \mathbf{v}(s)^\top \mathbf{r}_i \right]^2 \nonumber\\
&+ \mu \sum_{s=0}^{T-1} \|\mathbf{v}(s)\|^2
+ \gamma \sum_{i=0}^{N-1} \|\mathbf{r}_i\|^2 \nonumber\\
\text{subject to} \quad &
\|\mathbf{r}_i\| = 1, \quad i = 0,\dots,N{-}1.
\label{eq:objective}
\end{align}
Here, \(\mu\) and \(\gamma\) penalize excessive magnitudes in velocities and directions to mitigate overfitting and enhance numerical stability. This non-convex problem is solved via an alternating minimization scheme detailed in Algorithm~\ref{alg:alt-opt-3d-velocity}.

\begin{algorithm}[!b]
\caption{Alternating Optimization for DoRF Construction}
\label{alg:alt-opt-3d-velocity}
\begin{algorithmic}[1]
\setlength{\abovedisplayskip}{0pt}
\setlength{\belowdisplayskip}{0pt}
\setlength{\abovedisplayshortskip}{0pt}
\setlength{\belowdisplayshortskip}{2pt}
\REQUIRE Doppler projections \(\mathbf{V}_r\in\mathbb{R}^{T\times N}\), tolerance \(\epsilon\), regularizers \(\mu,\gamma, \lambda\)
\STATE Initialize \(\mathbf{R}=[\mathbf{r}_1,\dots,\mathbf{r}_N]\in\mathbb{R}^{3\times N}\) with unit columns
\REPEAT
  \STATE \textbf{Velocity update:}
  \[
  \mathbf{V} \leftarrow \mathbf{V}_r \mathbf{R}^\top\!\left(\mathbf{R}\mathbf{R}^\top+\lambda \mathbf{I}_3\right)^{-1}\in\mathbb{R}^{T\times 3}
  \]
  \STATE \textbf{Direction update:}
  \[
  \widetilde{\mathbf{R}}\leftarrow\left(\mathbf{V}^\top \mathbf{V}+\gamma \mathbf{I}_3\right)^{-1}\!\mathbf{V}^\top \mathbf{V}_r\in\mathbb{R}^{3\times N}
  \]
  \[
  \mathbf{r}_i\leftarrow \widetilde{\mathbf{r}}_i/\|\widetilde{\mathbf{r}}_i\|\quad\forall i
  \]
  \STATE \textbf{Prediction:}
  \[
  \widehat{\mathbf{V}}_r\leftarrow \mathbf{V} \mathbf{R}
  \]
  \STATE \textbf{Loss:}
  \[
  \mathcal{L}=\tfrac{1}{TN}\|\widehat{\mathbf{V}}_r-\mathbf{V}_r\|_F^{2}+\tfrac{\mu}{T}\|\mathbf{V}\|_F^{2}+\tfrac{\gamma}{N}\|\mathbf{R}\|_F^{2}
  \]
\UNTIL{\(\mathcal{L}<\epsilon\) or maximum iterations}
\STATE \textbf{DoRF construction:} let \(\mathbf{D}\in\mathbb{R}^{3\times K}\) be a uniform grid on the unit sphere; compute
\[
\mathbf{P}\leftarrow \mathbf{V} \mathbf{D}\in\mathbb{R}^{T\times K}
\]
\RETURN \(\mathbf{P}\in\mathbb{R}^{T\times K}\), \(\mathbf{V}\in\mathbb{R}^{T\times 3}\), \(\mathbf{R}\in\mathbb{R}^{3\times N}\)
\end{algorithmic}
\end{algorithm}

While the recovered velocity sequence \(\mathbf{V}\) provides a compact summary of motion, it remains frame dependent and lacks uniform directional coverage. To obtain viewpoint invariance, the velocities are re-projected onto an approximately uniform set of unit vectors on the sphere. Let \(\{\mathbf{d}_k\}_{k=1}^{K}\subset\mathbb{S}^2\) denote these directions. The radial component along \(\mathbf{d}_k\) is
\[
\mathbf{P}(s,k)=\mathbf{v}(s)^\top \mathbf{d}_k,\qquad k=1,\dots,K,
\]
which forms the DoRF matrix \(\mathbf{P}\in\mathbb{R}^{T\times K}\). Aggregating projections over \(\{\mathbf{d}_k\}\) yields an ordered, robust motion descriptor suitable for downstream HAR models.

\section{Method}

This section describes CSI phase pre-processing and a two-stage procedure that mitigates the impact of CSI noise in DoRF-based Wi-Fi HAR by using DoRF convergence behavior to automatically select reliable antennas.

\subsection{CSI Pre-processing}

Hardware imperfections such as sampling-frequency and symbol-timing offsets (SFO, STO) inject linear phase errors into raw CSI, degrading activity recognition. Following \cite{tadayon2019decimeter}, in this work, the phase is sanitized by unwrapping across subcarriers and subtracting a fitted linear trend that models the dominant hardware bias. The cleaned CSI is then used for delay–Doppler transformation, DoRF construction, and antenna selection.

\subsection{DoRF-Guided Antenna Selection}
\label{sec:dorf-antenna-selection}

Let APs be indexed by \(q\in\{1,\dots,Q\}\). AP \(q\) has antennas \(\mathcal{A}_q=\{1,\dots,M_q\}\) and Doppler projections \(\mathbf{V}_r^{(q)}\in\mathbb{R}^{T\times N_q}\). Columns of \(\mathbf{V}_r^{(q)}\) (delay bins/paths) are associated to antennas via a known mapping \(g_q:\{1,\dots,N_q\}\!\to\!\mathcal{A}_q\). Applying Algorithm~\ref{alg:alt-opt-3d-velocity} to each AP yields \(\mathbf{V}^{(q)}\in\mathbb{R}^{T\times 3}\), \(\mathbf{R}^{(q)}\in\mathbb{R}^{3\times N_q}\), and \(\widehat{\mathbf{V}}_r^{(q)}=\mathbf{V}^{(q)}\mathbf{R}^{(q)}\). For antenna \(a\in\mathcal{A}_q\) with index set \(\mathcal{I}_{q,a}=\{i:\,g_q(i)=a\}\), define the (normalized) per-antenna DoRF convergence error
\begin{equation}
\label{eq:antenna-error}
\mathcal{E}_{q,a}
=\frac{\displaystyle\sum_{i\in\mathcal{I}_{q,a}}\!\!\big\|\mathbf{V}_r^{(q)}(:,i)-\mathbf{V}^{(q)}\mathbf{r}^{(q)}_i\big\|_2^2}
{\,\displaystyle\sum_{i\in\mathcal{I}_{q,a}}\!\!\big\|\mathbf{V}_r^{(q)}(:,i)\big\|_2^2+\delta\,},\qquad \delta>0,
\end{equation}
where \(\mathbf{r}^{(q)}_i\) is the \(i\)-th column of \(\mathbf{R}^{(q)}\). Collect all antenna errors over all APs into \(\{\mathcal{E}_{q,a}\}\) with the total count \(M_Q=\sum_{q}M_q\). Sorting them in nondecreasing order gives \(\tilde{e}_{(1)}\le\cdots\le\tilde{e}_{(M_Q)}\). A knee index \(k^\star\) is found by the farthest-point (triangle) method on \((i,\tilde{e}_{(i)})\): let \(A=(1,\tilde{e}_{(1)})\), \(B=(M_Q,\tilde{e}_{(M_Q)})\), \(u=B\!-\!A\). The perpendicular distance of \(X_i=(i,\tilde{e}_{(i)})\) to line \(AB\) is
\begin{equation}
\label{eq:knee-distance}
d_i=\frac{\big|u_x(\tilde{e}_{(i)}-\tilde{e}_{(1)})-u_y(i-1)\big|}{\sqrt{u_x^2+u_y^2}},\quad
u=(u_x,u_y),
\end{equation}
and \(k^\star=\arg\max_i d_i\). The globally selected antenna set is
\begin{equation}
\label{eq:selected-set}
\mathcal{S}=\big\{(q,a):\, \mathcal{E}_{q,a}\le \tilde{e}_{(k^\star)}\big\},
\end{equation}
which adapts to varying noise levels and environmental conditions without per-AP thresholds.

\begin{algorithm}[!b]
\caption{DoRF-Guided Antenna Selection and Re-fitting}
\label{alg:dorf-antenna-selection}
\begin{algorithmic}[1]
\setlength{\abovedisplayskip}{0pt}
\setlength{\belowdisplayskip}{0pt}
\setlength{\abovedisplayshortskip}{0pt}
\setlength{\belowdisplayshortskip}{2pt}
\REQUIRE For each AP \(q\): Doppler projections \(\mathbf{V}_r^{(q)}\), mapping \(g_q\); small \(\delta>0\)
\vspace{4pt}
\STATE \textbf{Stage 1: Global antenna selection}
\STATE \hspace{1.0em}\textbf{(a) Per-AP DoRF fitting:} For each \(q\), run Algorithm~\ref{alg:alt-opt-3d-velocity} on \(\mathbf{V}_r^{(q)}\) to obtain \(\mathbf{V}^{(q)},\mathbf{R}^{(q)}\) and \(\widehat{\mathbf{V}}_r^{(q)}=\mathbf{V}^{(q)}\mathbf{R}^{(q)}\).
\STATE \hspace{1.0em}\textbf{(b) Per-antenna errors:} For each antenna \(a\in\mathcal{A}_q\), compute \(\mathcal{E}_{q,a}\) via \eqref{eq:antenna-error}.
\STATE \hspace{1.0em}\textbf{(c) Pool \& rank across all APs:} Form the multiset \(\{\mathcal{E}_{q,a}\}\), sort to obtain \(\tilde{e}_{(i)}\), compute distances \(d_i\) by \eqref{eq:knee-distance}.
\STATE \hspace{1.0em}\textbf{(d) Knee threshold \& selection:} Let \(\tau=\tilde{e}_{(k^\star)}\), set \(\mathcal{S}=\{(q,a): \mathcal{E}_{q,a}\le\tau\}\).
\vspace{6pt}
\STATE \textbf{Stage 2: Per-antenna DoRF re-fitting (on selected antennas)}
\STATE \hspace{1.0em}\textbf{(e) Re-fit DoRFs:} For each \((q,a)\in\mathcal{S}\), run Algorithm~\ref{alg:alt-opt-3d-velocity} on \(\mathbf{V}_r^{(q)}(:,\mathcal{I}_{q,a})\) to obtain a per-antenna DoRF \(\mathbf{P}^{(q,a)}\in\mathbb{R}^{T\times K}\).
\STATE \hspace{1.0em}\textbf{(f) Export features:} Store or aggregate \(\{\mathbf{P}^{(q,a)}\}_{(q,a)\in\mathcal{S}}\) for downstream classification.
\RETURN Selected set \(\mathcal{S}\) and per-antenna DoRFs \(\{\mathbf{P}^{(q,a)}\}\).
\end{algorithmic}
\end{algorithm}

\subsection{Per-Antenna DoRF Fitting and Activity Classification}

The structured DoRF \(\mathbf{P}\in\mathbb{R}^{T\times M\times 2M}\), obtained by uniform spherical sampling (\(2M^2\) directions), serves as the input to MORIC~\cite{hasanzadeh2025moric}. Each directional Doppler projection is independently encoded by a feature extractor that applies random convolutional kernels followed by temporal pooling, yielding vectors \(\mathbf{f}_{s,i}\in\mathbb{R}^d\) for \(i=1,\dots,2M^2\). As Doppler directions on the fitted DoRF spheres can differ in order and rotation, MORIC attains invariance by applying element-wise max-pooling along the projection axis,
\[
\mathbf{f}_s=\max_{i=1,\dots,2M^2} \mathbf{f}_{s,i}.
\]

MORIC then passes the pooled feature vector through shallow fully connected layers with non-linear activations to produce activity class probabilities.

\section{Experiment}
\label{sec:pagestyle}
\subsection{Data}

\begin{figure}[!t]
	\centering
	\subfloat{%
		\includegraphics[width=0.75\linewidth]{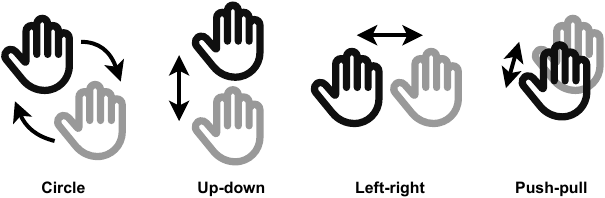}
	}
	\caption{Illustration of the four hand gestures performed by six participants in the UTHAMO dataset.
}
\label{figure:gestures}
\end{figure}

\begin{figure}[!t]
        \centering		\includegraphics[width=0.66\linewidth]{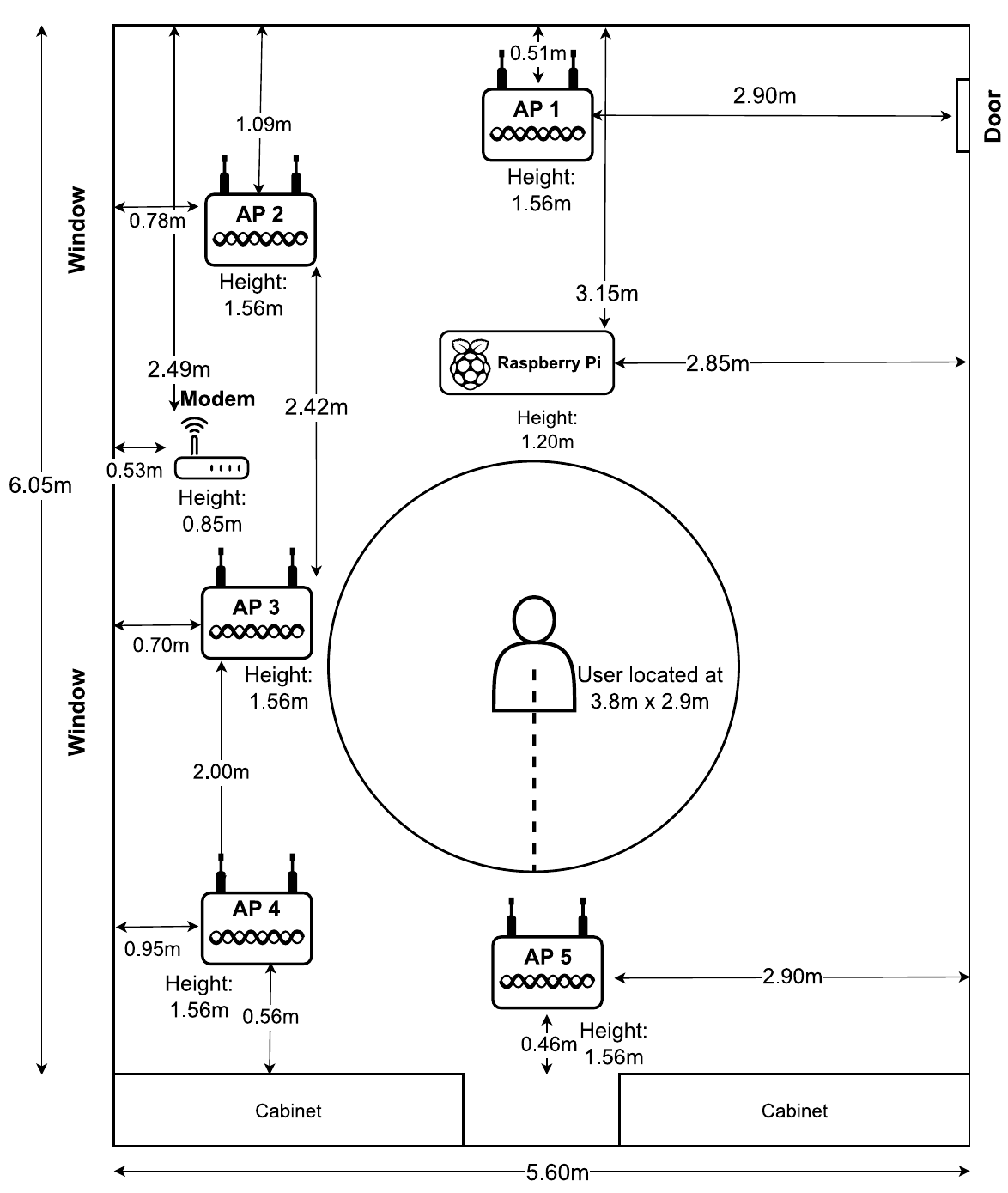}
	
	\caption{The floor plan of the UTHAMO data collection setup.}
	\vspace{-0mm}
\label{figure:map}
\end{figure}

The UTHAMO dataset \cite{uthamo} is used to evaluate the method. Six participants performed four gestures—\textit{circle}, \textit{left–right}, \textit{up–down}, and \textit{push–pull}—as illustrated in Fig.~\ref{figure:gestures}. Data were collected in a static indoor office (\(6\,\mathrm{m}\times 5.6\,\mathrm{m}\)); the floor plan is shown in Fig.~\ref{figure:map}. CSI was recorded at \(2.4\,\mathrm{GHz}\) using the Nexmon toolkit \cite{nexmon}, between a Raspberry Pi transmitter and five ASUS RT-AC86U Wi-Fi routers (three antennas each) positioned around the user, yielding \(64\) subcarriers per antenna. Each gesture comprised \(20\) trials, each lasting \(5\,\mathrm{s}\), sampled at \(100\,\mathrm{Hz}\).

\subsection{Training and Test Procedure}

The default hyperparameters from \cite{hasanzadeh2025moric} were used for the MORIC architecture. It employed \(K=2\) attention heads, \(D=1{,}000\) random kernels, a hidden size of \(256\), and a reduced feature dimension \(D^{\prime}=128\). Training used the AdamW optimizer~\cite{loshchilov2017decoupled} (learning rate \(1\times 10^{-4}\), batch size \(64\)) with the cross-entropy loss and label smoothing \(\alpha=0.1\). A maximum of \(2{,}500\) epochs was used, with early stopping after \(200\) epochs without validation improvement. The DoRF was constructed per antenna and merged to form the MORIC input, using a uniform grid \(M=8\) and an early stopping threshold \(\varepsilon=0.01\) in the alternating optimization. Generalization was assessed using leave-one-subject-out (LOSO) cross-validation, selecting the model with the lowest validation loss for the final evaluation. The DoRFs of discarded antennas are replaced with zero vectors to maintain a consistent input size.

\subsection{Results}
\label{sec:typestyle}
\begin{figure}[!b]
	\centering
	\subfloat{%
		\includegraphics[width=0.80\linewidth]{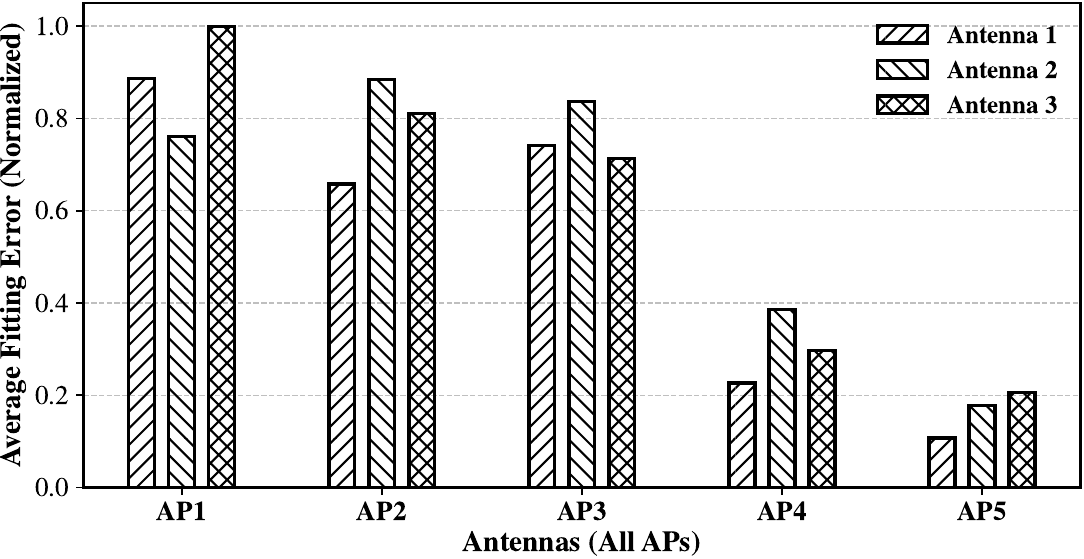}
	}
	\caption{Per-AP DoRF fitting error across antennas, averaged over all samples.
}
\label{figure:hist}
\end{figure}

Table~\ref{table:results} summarizes unseen user generalization for four class HAR on the UTHAMO dataset. Magnitude-only CSI baselines AMAP and CMAP perform near chance, while Doppler and phase aware methods such as CapsHAR and the CSI Ratio provide only modest gains from \(33.7\%\) to \(36.1\%\). MORIC, which uses Doppler velocity projections, raises the mean accuracy to \(56.3\%\), and the construction of DoRFs increases it to \(58.9\%\). With an accuracy of \(\mathbf{65.3\%} \pm \mathbf{8.3\%}\), the proposed DoRF-guided antenna selection followed by DoRF re-fitting outperforms all state-of-the-art methods, demonstrating its effectiveness in preserving antennas that yield reliable and consistent motion information.

Fig.~\ref{figure:hist} shows the average DoRF fitting error across antennas and APs in the UTHAMO setup. AP$5$ and AP$4$ exhibit the lowest errors and capture more consistent Doppler projections. The remaining APs have higher errors, indicating greater noise. This pattern reflects the deployment geometry, as gestures were performed in front of AP$5$ and AP$4$ along their line-of-sight to the Raspberry Pi transmitter, while the other APs did not intersect with the users' movements. These results highlight the potential of the proposed method to adaptively disregard contributions from APs that capture little motion-related information.

\begin{table}[!b]
\caption{Generalization results for four-class hand motion recognition using multiple Wi-Fi APs, averaged over six users.}
\small % reduce font size
\setlength{\tabcolsep}{5pt} % reduce horizontal padding
\begin{tabular*}{\linewidth}{@{\extracolsep{\fill}}lcc}
  \toprule
  \multirow{2}{*}{Method} &
    \multicolumn{2}{c}{Accuracy (\%)} \\
    & {Mean} & {Standard Deviation} \\
  \midrule
  AMAP \cite{salehinejad2023joint} & $26.3$ & $2.8$ \\
  CMAP \cite{salehinejad2023joint} & $28.1$ & $2.9$ \\
  CapsHAR \cite{djogoHAR} & $33.7$ & $6.1$ \\    
  CSI Ratio Model \cite{wu2022wifi} & $36.1$ & $5.1$ \\
  MORIC \cite{hasanzadeh2025moric} & $56.3$ & $9.1$ \\
    DoRF \cite{hasanzadeh2025dorf} & $58.9$ & $8.1$ \\
  \hline
  \textbf{Proposed Method (DoRF + Ant. Sel.)} & $\mathbf{65.3}$ & $\mathbf{8.3}$ \\
  \bottomrule
\end{tabular*}
\label{table:results}
\end{table}

\section{Conclusion}\label{section:conclusion}
This work introduces a novel DoRF-guided, Wi-Fi–based HAR framework for APs with multiple antennas, targeting the challenging task of small-scale hand gesture recognition, where motion signals are subtle and highly susceptible to noise. The method employs each antenna’s DoRF fitting error to quantify motion-related Doppler information and cross-antenna consistency, discarding antennas that provide weak or inconsistent cues. By retaining only the most informative antennas, it preserves critical motion dynamics and delivers a robust motion representation. Experiments demonstrate substantial improvements in cross-user generalization, establishing the proposed framework as a robust solution for fine-grained Wi-Fi sensing.

\newpage

\bibliographystyle{IEEEbib}
\bibliography{refs}

\label{sec:refs}

\end{document}